\documentclass[letterpaper]{article}
\usepackage{aaai17}
\usepackage{times}
\usepackage{helvet}
\usepackage{courier}
\usepackage{url}
\usepackage{graphicx}
\usepackage{amsmath}	% mathematics
\usepackage{mathtools}	% mathematics
\usepackage{xfrac}		% \sfrac
\usepackage{amssymb}	% math symbols
\usepackage{booktabs}	% nicer tables
\usepackage{siunitx}	% using \num
\usepackage{xspace}		% fix space in macros
\usepackage[font={small}]{subfig} % subfig, 4 figures in a row
\begin{document}
\setlength{\titlebox}{2.5in}
\pdfinfo{
/Title (The Ebb and Flow of Controversial Debates on Social Media)
/Author (Kiran Garimella, Gianmarco De Francisci Morales, Aristides Gionis, Michael Mathioudakis)
/Keywords (Controversy, Polarization, Social Media, Twitter)
}
%
%Section Numbers
% Uncomment if you want to use section numbers
% and change the 0 to a 1 or 2
\setcounter{secnumdepth}{1}
%
% Title and Author Information Must Immediate Follow
% the pdfinfo within the preamble
%
\title{The Ebb and Flow of Controversial Debates on Social Media}
\author{Kiran Garimella\textsuperscript{1}\\
Aalto University\\
Helsinki, Finland\\
{kiran.garimella@aalto.fi}\\
\And Gianmarco~De~Francisci~Morales\\
QCRI\\
Doha, Qatar\\
{gdfm@acm.org}\\
\AND Aristides Gionis\\
Aalto University\\
Helsinki, Finland\\
{aristides.gionis@aalto.fi}\\
\And Michael Mathioudakis\textsuperscript{1}\\
Aalto University \& HIIT\\
Helsinki, Finland\\
{michael.mathioudakis@aalto.fi}\\
% \AND\\
{\scriptsize{\textsuperscript{1}Contact authors; contributed equally.}}
}

% Our macros
% Names and symbols
\newcommand{\rwc}{{{\sc RWC}}\xspace}
% Paragraphs
\newcommand{\spara}[1]{\smallskip\noindent\textbf{#1}}
\newcommand{\mpara}[1]{\medskip\noindent\textbf{#1}}
\newcommand{\para}[1]{\noindent\textbf{#1}}

% TO REMOVE %% TODO
\newcommand{\np}{\ensuremath{\mathbf{NP}}}
\newcommand{\nphard}{{{\np}-hard}\xspace}
\newcommand{\kedgeadd}{{\ensuremath{k}}-{\sc Edge\-Addition}\xspace}
\newcommand{\kedgeaddexp}{{\ensuremath{k}}-{\sc Edge\-Addition\-Expectation}\xspace}
\newcommand{\kedgeaddsimple}{{\ensuremath{k}}-{\sc Edge\-Addition}{\ensuremath{_{0}}}\xspace}
\newcommand{\pagerank}[2]{{\ensuremath{\mathrm{pr}_{#1}({#2})}}\xspace}
\newcommand{\vertexcover}{{\sc Vertex\-Cover}\xspace}
\newcommand{\metis}{{\sc metis}\xspace}
\newcommand{\rov}{\ensuremath{\text{\sc ROV}}\xspace}
\newcommand{\rovap}{\ensuremath{\text{\sc ROV-AP}}\xspace}
\newcommand{\core}{c\xspace}
\newcommand{\side}{s\xspace}
\newcommand{\words}{w\xspace}

\maketitle
\begin{abstract}
We explore how the polarization around 
controversial topics evolves on Twitter --
over a long period of time (2011 to 2016), and also
as a response to major external events that lead to
increased related activity. We find that increased
activity is typically associated with increased polarization;
however, we find no consistent long-term trend in polarization
over time among the topics we study.
\end{abstract}

\section{Introduction}
\enlargethispage{\baselineskip}

% this work
We study how online discussions around controversial topics change as interest in them increases and decreases -- or ``ebbs and flows''.
We are motivated by the observation that interest in enduring controversial 
issues is re-kindled by major related events.
The gun control debate in U.S., which is revived whenever a mass shooting occurs, is one such example.
The occurrence of such an event commonly causes an increase in collective attention, as reflected in the volume of related activity in social media.
Moreover, motivated by the common perception that political polarization has risen recently, we track the polarization of controversial topics over the span of multiple years.

% data
Specifically, our study is based on  Twitter data
that cover five years (2011 to 2016).
We track four popular controversial topics of discussion in the U.S. that are recurring and attracted considerable attention 
during the 2016 U.S. election cycle.
Given a controversial topic, we build a \emph{endorsement} network from the retweet information on Twitter for each day of activity, and thus obtain a time series of endorsement networks. 
Following~\cite{garimella2016quantifying}, we measure the polarization reflected in each network instance by using the \texttt{Random Walk Controversy} (\rwc) measure.

Our analysis then seeks to answer two questions: firstly, how polarization varies with increased volume of activity (i.e., the number of users who actively discuss the topic) in a given day; and secondly, how polarization of each topic evolves over a long period of time.
With respect to the former, our findings suggest that increased volume of activity is associated with increased polarization.
% We supplement our analysis on controversial topics with an analysis on non-controversial ones. %% TODO do we include?
With respect to the latter, our findings suggest that there is no consistent long-term trend over time.

\section{Related Work}
\label{sec:related}

A few studies exist on the topic of controversy in online news and social media. 
In one of the first papers, \cite{adamic2005political}
study linking patterns and topic coverage of political bloggers, 
focusing on blog posts on the U.S.\ presidential election of 2004.
They measure the degree of interaction between liberal and conservative blogs,
and provide evidence that  conservative blogs are linking to each other more frequently and in a denser pattern.

These findings are confirmed by a more recent study of \cite{conover2011political}, 
who focus on political communication regarding congressional midterm elections.
Using data from Twitter, 
they identify a highly segregated partisan structure 
(evidenced in retweets but not replies), 
with limited connectivity between left- and right-leaning users.

In another recent work,  
\cite{mejova2014controversy} consider discussions of controversial and non-controversial news over a span of $7$ months.
They find a significant correlation between controversial issues and the use of negative affect and biased language. 
%polarized discussions have a certain structure and

More recently, \cite{garimella2016quantifying,garimella2016exploring} show that controversial discussions on social media have a well-defined structure, when looking at the \emph{endorsement} network.
They propose a measure based on random walks (\rwc), 
which is able to identify controversial topics, and 
\emph{quantify} the level of controversy of a given discussion on social media
by the structure of its endorsement network.

Unlike the aforementioned works, here we are interested in dynamic aspects of polarized networks. Previous studies with similar focus includes~\cite{Lehmann2012}, which examines spikes in the frequency of hashtags and identifies a classification scheme that predicts whether the hashtags correspond to endogenously or exogenously driven topics.
Other related studies works include~\cite{morales2015measuring}, who study polarization over time for the death of Hugo Chavez, and~\cite{andris2015rise}, who study the partisanship of the U.S.\ congress over a long period of time.

Moreover, we are particularly interested in the network response to external stimuli that lead to increased collective attention in the controversial topic -- an issue that only very recently has seen some attention in the literature~\cite{romero2016social}.

\section{Dataset}
\label{sec:dataset}

% Our study uses data collected from Twitter.
Using the repositories of the Internet Archive,\footnote{\small\url{https://archive.org/details/twitterstream}}
we collect a $1\%$ sample of tweets
from September 2011 to August 2016,\footnote{\small{To be precise, we have data for $57$ months from that period, as our data source has no data available for three months.}}
for four topics of discussion related to `Obamacare', `Abortion', `Gun Control', and `Fracking'.
These topics constitute long-standing controversial issues in the U.S.\footnote{\small{According to \small\url{http://2016election.procon.org}.}} and have been used in previous work~\cite{lu2015biaswatch}.
For each topic, we use a keyword list as proposed by~\cite{lu2015biaswatch} (shown in Table~\ref{tab:keywords}), and extract a base set of tweets containing at least one topic-related keyword.
To enrich this original dataset, we use the Twitter REST API to obtain all tweets of users who have participated in the discussion at least once.\footnote{\small Up to \num{3200} due to limits in the API.}
Table~\ref{tab:keywords} shows the final statistics for the dataset.

\begin{table}[t]
\centering
\small
\caption{Keywords for the controversial topics.}
\label{tab:keywords}
\begin{tabular}{l>{\raggedright}p{10em}rr}
\toprule
Topic & Keywords & \#Tweets & \#Users \\
\midrule
Obamacare & obamacare, \#aca & \num{866484} & \num{148571} \smallskip\\
Abortion & abortion, prolife, prochoice, anti-abortion, pro-abortion, planned parenthood & \num{1571363} & \num{327702} \smallskip\\
Gun Control & gun control, gun right, pro gun, anti gun, gun free, gun law, gun safety, gun violence& \num{824364} & \num{224270} \smallskip\\
Fracking & fracking, \#frack, hydraulic fracturing, shale, horizontal drilling & \num{2117945} & \num{170835} \\ 
\bottomrule
\end{tabular}
\vspace{-1.1\baselineskip}
\end{table}

% networks
We process the dataset to build {\em retweet networks} --- i.e., directed networks of users, where there is an edge between two users ($u\,{\rightarrow}\,v$) if $u$ retweets $v$.
% sides
Polarized networks, particularly the ones considered here, can be broadly characterized by two opposing \emph{sides}, which express different opinions on the associated topic.
It is commonly understood that retweets indicate \emph{endorsement}, and endorsement networks for controversial topics have been shown to have a bi-clustered structure~\cite{conover2011political,garimella2016quantifying,garimella2017connecting}, i.e., they consist of two well-separated clusters that correspond to the opposing points of view on the topic.

% There is evidence that retweet networks are capable of identifying the different sides of the controversy~\cite{conover2011political}.
In this paper, we build upon this observation to reveal the opposing sides around a topic.
In particular, following an approach from previous work~\cite{garimella2016quantifying}, we collapse all retweets contained in the dataset of each topic into a single large static retweet network --- and use the METIS clustering algorithm~\cite{karypis1995metis} to identify two clusters that correspond to the two opposing sides.
% This process allows us to identify consistent sides for the topic.
We evaluate the sides by manual inspection of the top retweeted users, URLs, and hashtags.
The results are fairly consistent and accurate, and can be inspected online.\footnote{\small{https://mmathioudakis.github.io/polarization/}\label{footnote:website}}

\begin{figure*}[t]
\centering
\includegraphics[width=\textwidth]{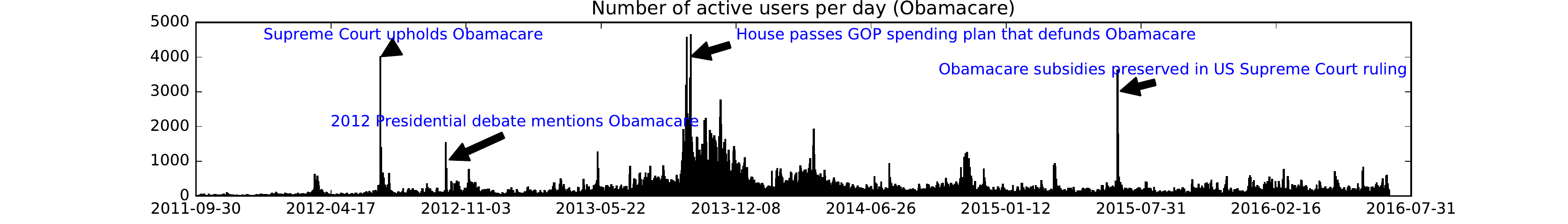}
\includegraphics[width=\textwidth]{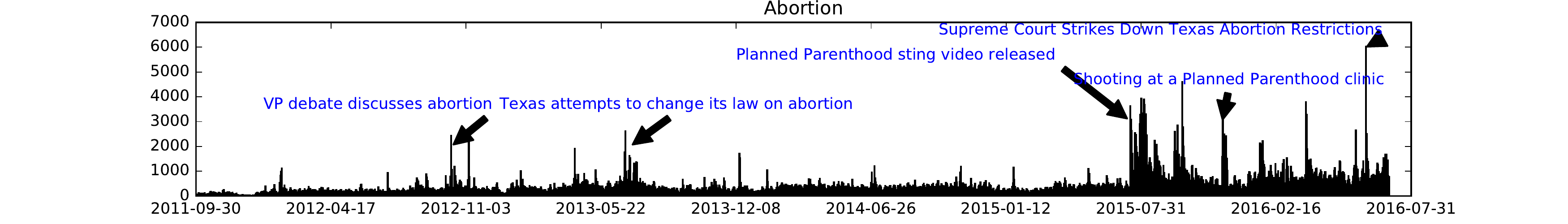}
\includegraphics[width=\textwidth]{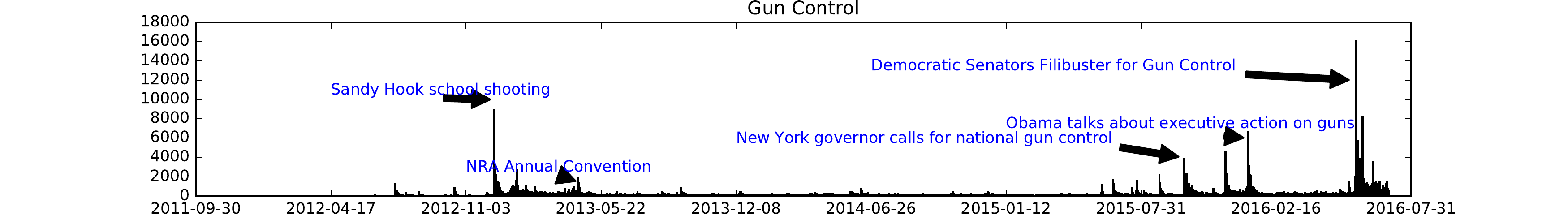}
\includegraphics[width=\textwidth]{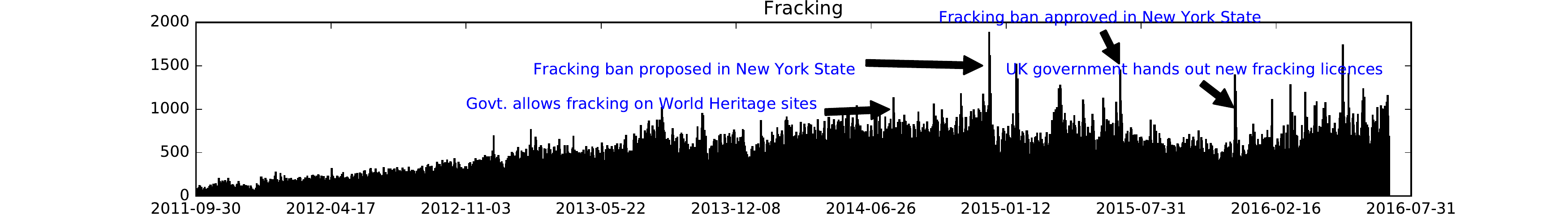}
\caption{Daily trends for number of active users for the four controversial topics under study. Clear spikes occur at several points in the timeline. Manually chosen labels describing related events reported in the news on the same day are shown in blue for some of the spikes.}
\label{fig:timeline}
\end{figure*}

% time granularity
% Let us now consider the temporal dynamics of these interaction networks.
Given the traditional daily news reporting cycle, we build a time series of retweet networks with the same granularity.
% This resolution allows us to identify spikes in the level of interest in a topic -- a method that allows to identify linked to real world external events~\cite{mathioudakis2010twittermonitor}.
At this granularity level, we can easily identify spikes of interest in the topics, as seen in Figure~\ref{fig:timeline}.
Such spikes commonly correspond to external newsworthy events~\cite{mathioudakis2010twittermonitor}.
Figure~\ref{fig:timeline} is manually annotated with major events associated with the observed spikes.
These results support the observation that Twitter is used as an \emph{agor\'{a}} to discuss the daily matters of public interest~\cite{deFrancisciMorales2012trex}.

\spara{Notation}
The set of retweets that occur within a single day $d$ gives rise to one retweet network $N^\mathit{rt}_d$.
Each user associated with a retweet is represented with one node in the network.
There is a directed edge from user $u$ to user $v$ only when user $u$ has retweeted at least one tweet authored by user $v$.
In addition, each node $u$ in the network is associated with a binary attribute $\side(u) \in \{1, 2\}$
that represents the side the node belongs to.

\section{Analysis}
\label{sec:analysis}

We quantify the polarization of a network $N_d$ via the random walk controversy (\rwc) score introduced in previous work~\cite{garimella2016quantifying}.
Intuitively, the score captures whether the network consists of two well-separated clusters.
The higher the value of the score, the clearer the separation between the two sides.

\subsection{Polarization vs Collective Attention}
We explore how \rwc\ varies with the number of users in the networks, which is a proxy for the amount of collective attention the topic attracts.
We sort the time series of networks by volume of active users, and partition them into ten quantiles (each having an equal number of days),
so that days of bucket $i$ are associated with smaller volume than those of bucket $j$, 
for $i < j$.
For each bucket, we report the mean and standard deviation of the values for each measure, and observe the trend from lower to higher volume.

Note that \rwc\ is carefully defined so that its expected value does not depend on the volume of underlying activity (i.e., number of network nodes).

\begin{figure*}[tb]
\centering
\begin{minipage}{.22\linewidth}
\centering
\subfloat{\label{}\includegraphics[width=\textwidth]{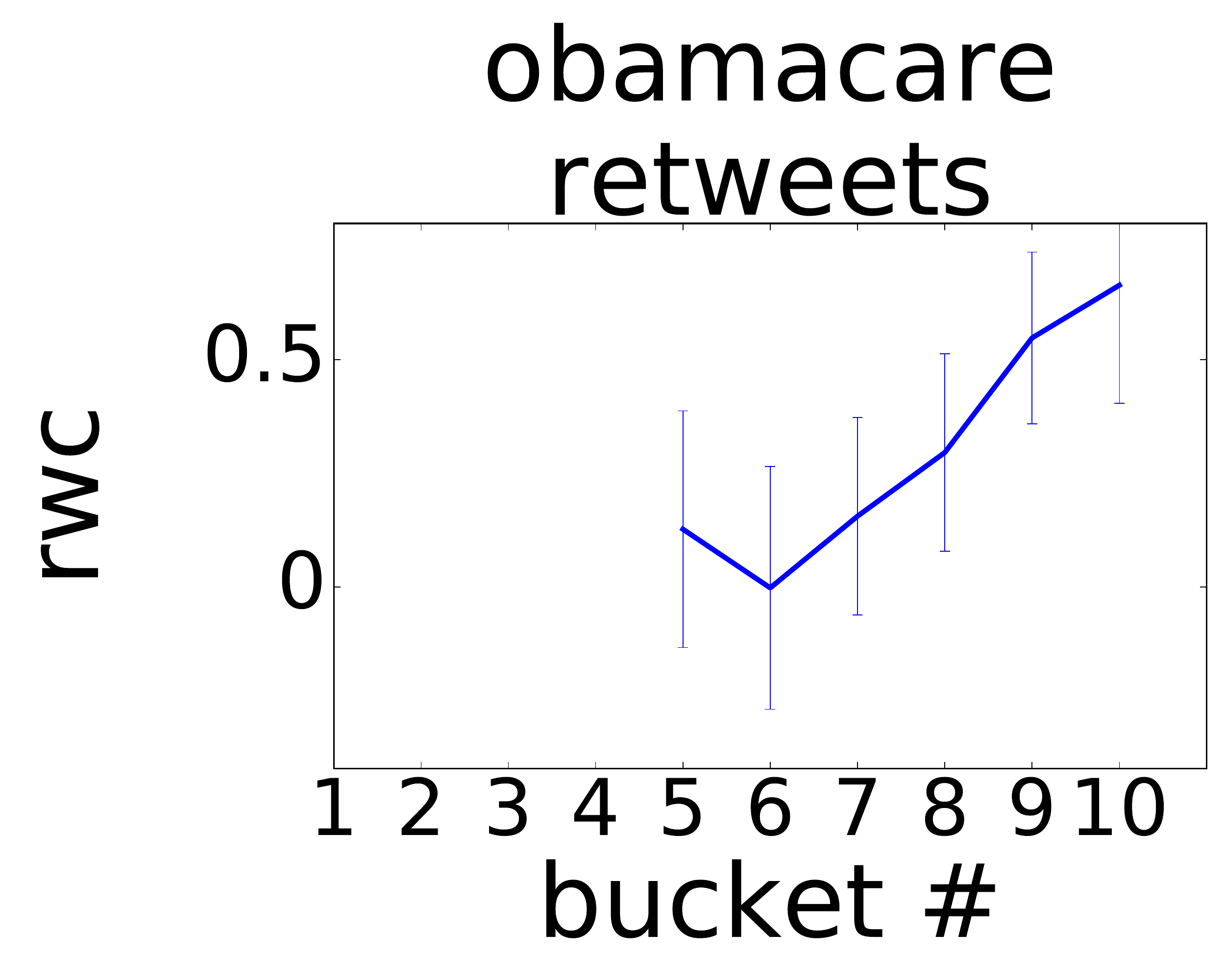}}
\end{minipage}%
\begin{minipage}{.22\linewidth}
\centering
\subfloat{\label{}\includegraphics[width=\textwidth]{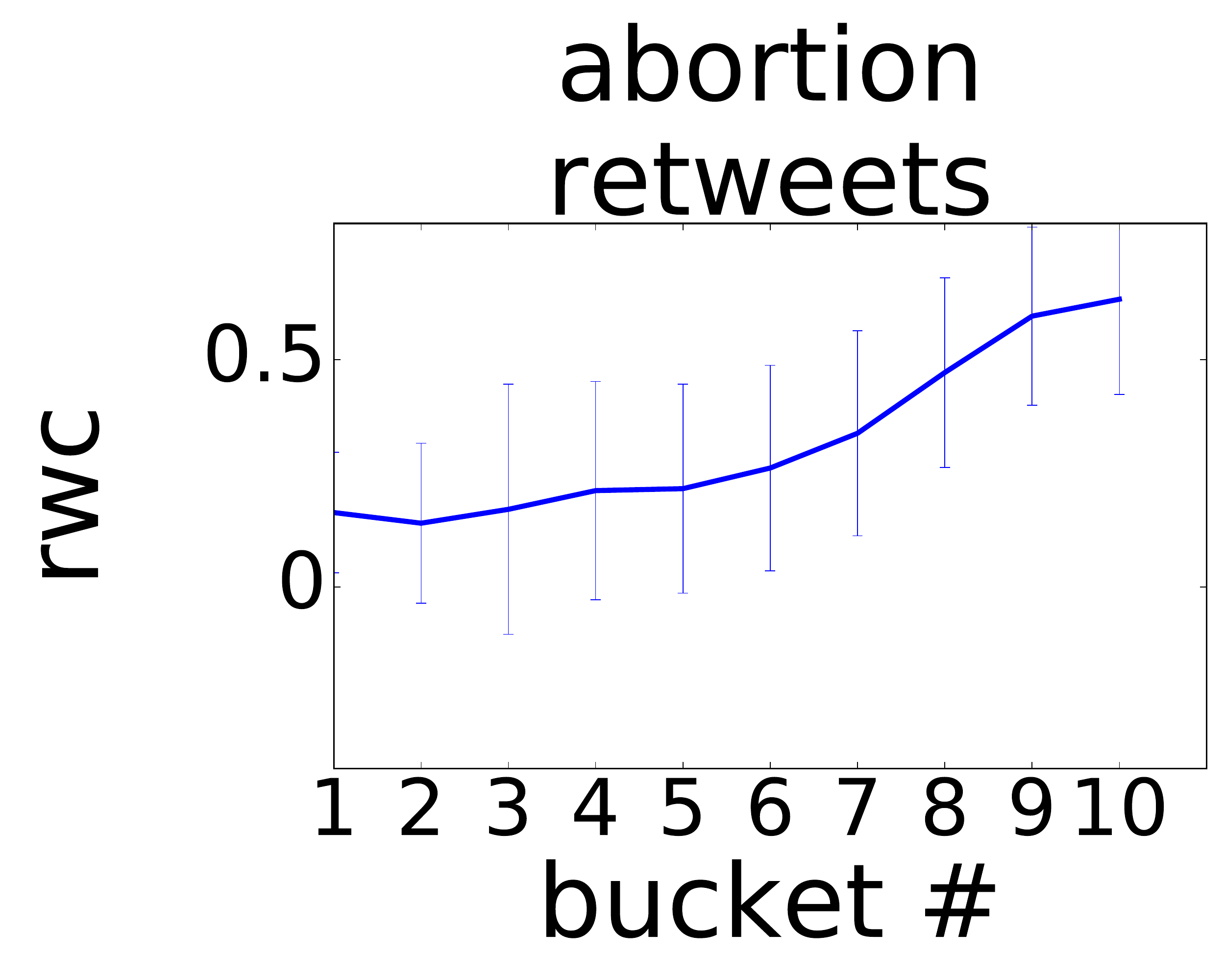}}
\end{minipage}%
\begin{minipage}{.22\linewidth}
\centering
\subfloat{\label{}\includegraphics[width=\textwidth]{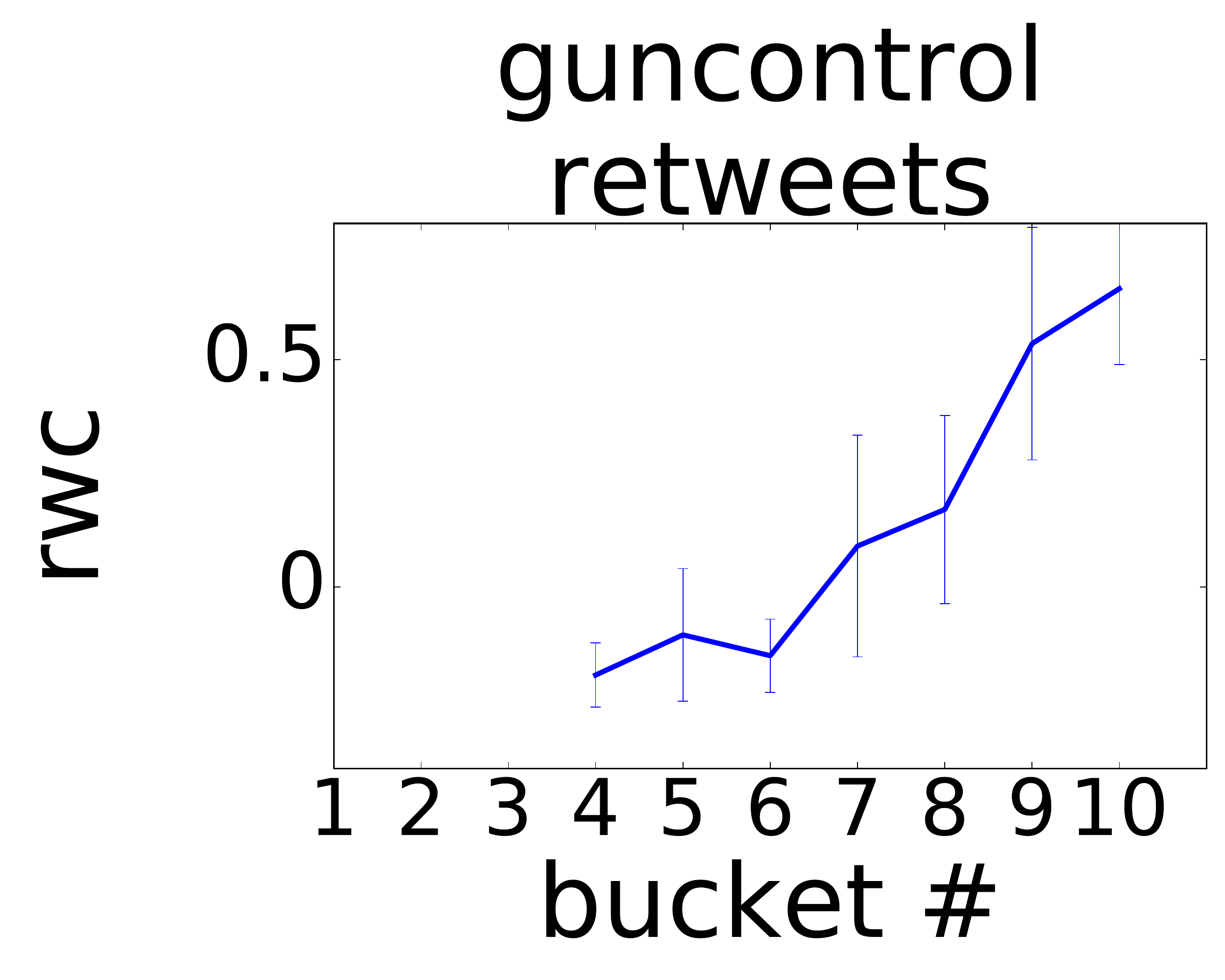}}
\end{minipage}%
\begin{minipage}{.22\linewidth}
\centering
\subfloat{\label{}\includegraphics[width=\textwidth]{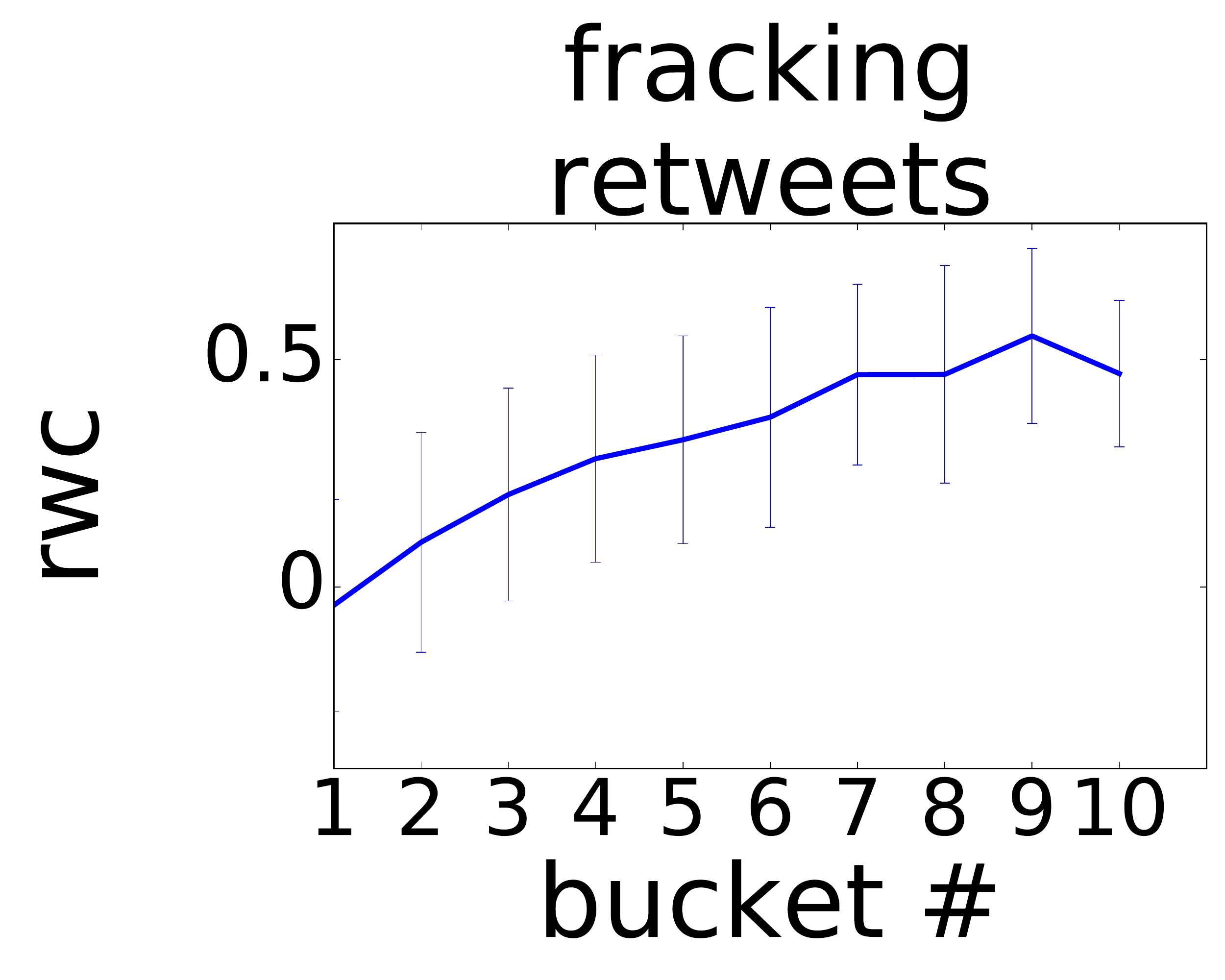}} % TODO
\end{minipage}%
\caption{RWC score as a function of the activity volume in the retweet network. An increase in interest in the controversial topic corresponds to an increase in the controversy score of the retweet network.}
\label{fig:rwc}
\end{figure*}

We observe a significant pattern in the relationship between \rwc and interest in the topic.
Figure~\ref{fig:rwc} shows \rwc as a function of the quantiles of the network by retweet volume.
There is a clear increasing trend, which is consistent across topics.
This trend suggests that increased interest in the topic is correlated with an increase in controversy of the debate, 
and increased polarization of the retweet networks for the two sides.

\subsubsection{Non-controversial topics.}

\begin{figure*}
\centering
\begin{minipage}{.22\linewidth}
\centering
\subfloat{\label{}\includegraphics[width=\textwidth]{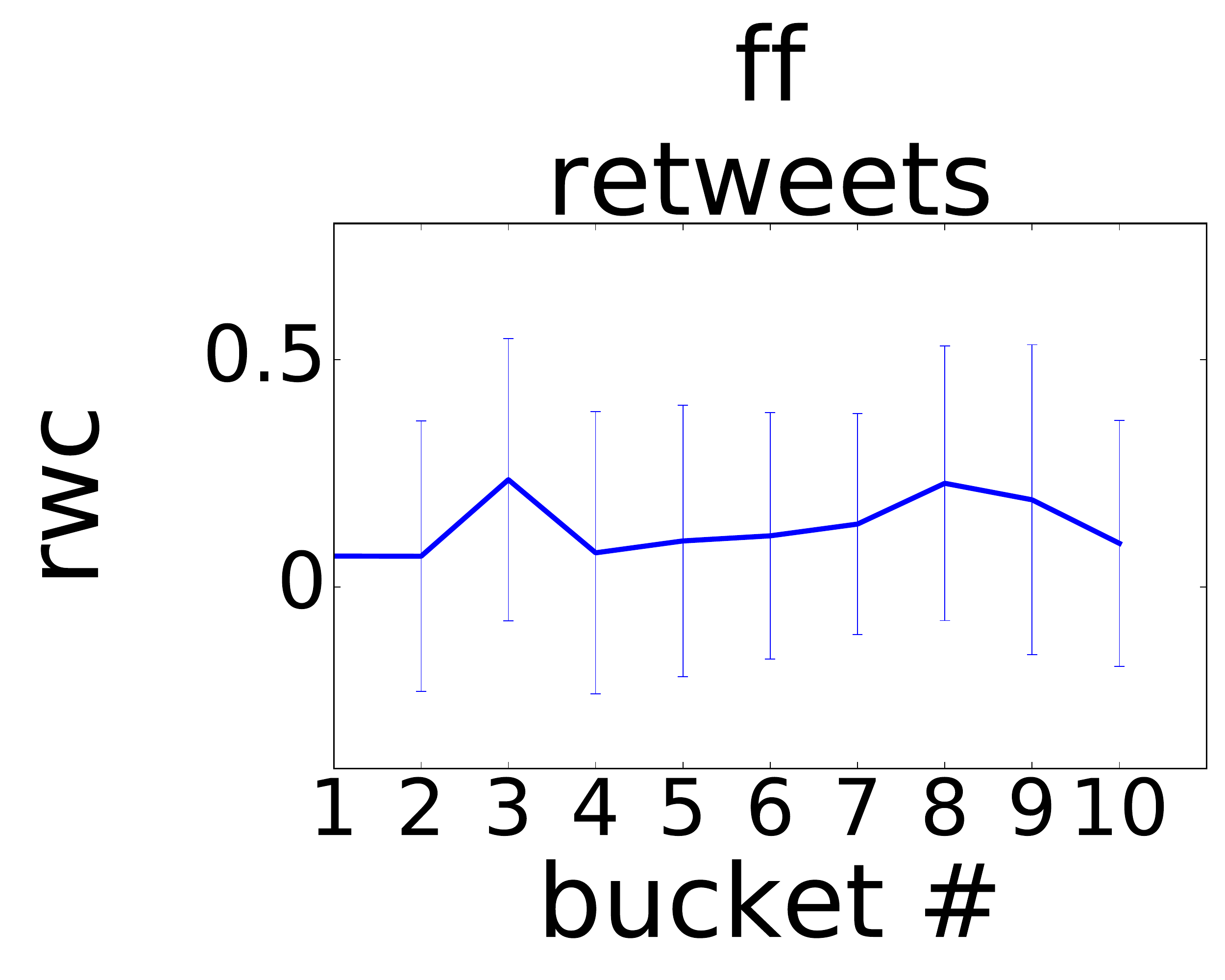}}
\end{minipage}%
\begin{minipage}{.22\linewidth}
\centering
\subfloat{\label{}\includegraphics[width=\textwidth]{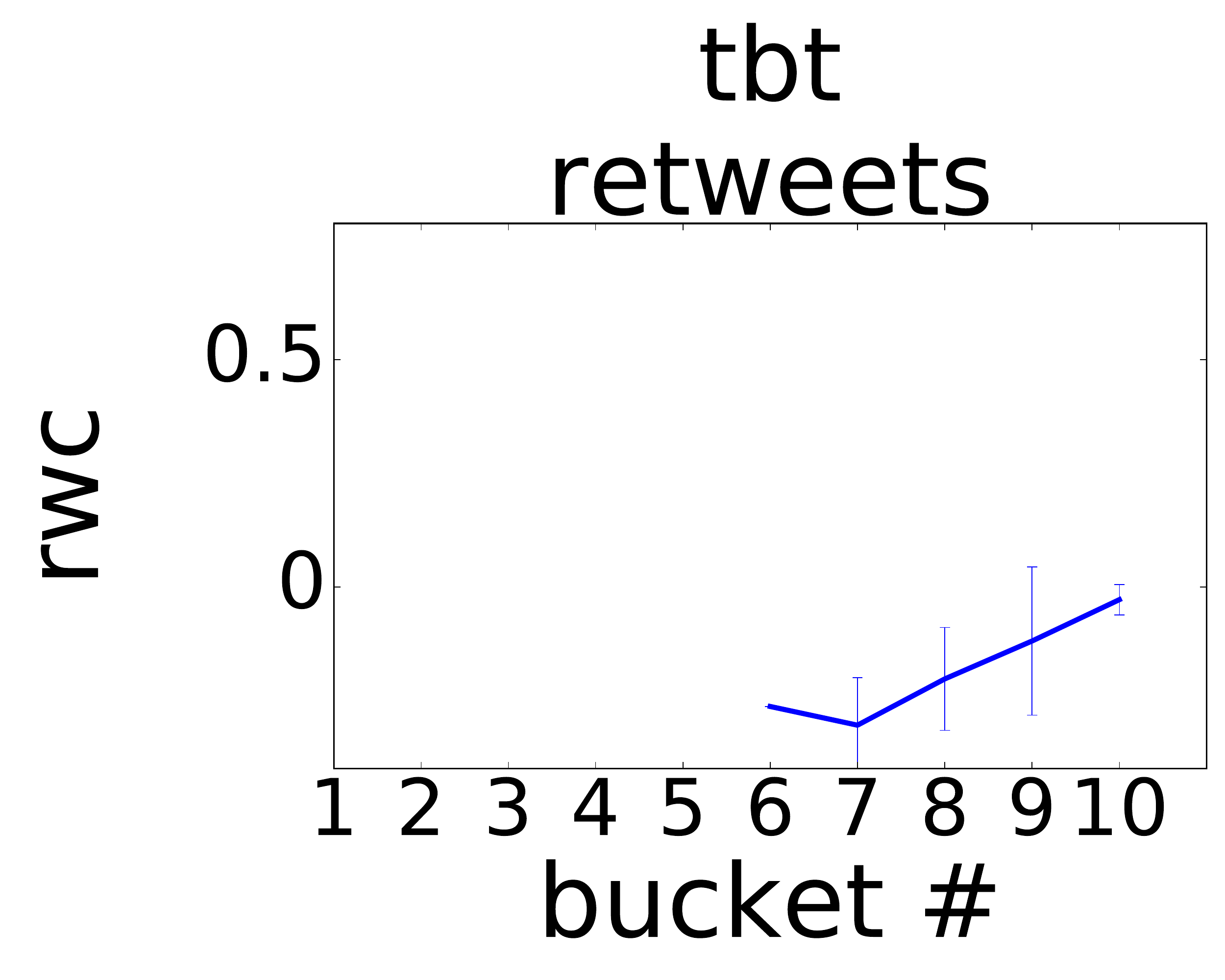}}
\end{minipage}%
\begin{minipage}{.22\linewidth}
\centering
\subfloat{\label{}\includegraphics[width=\textwidth]{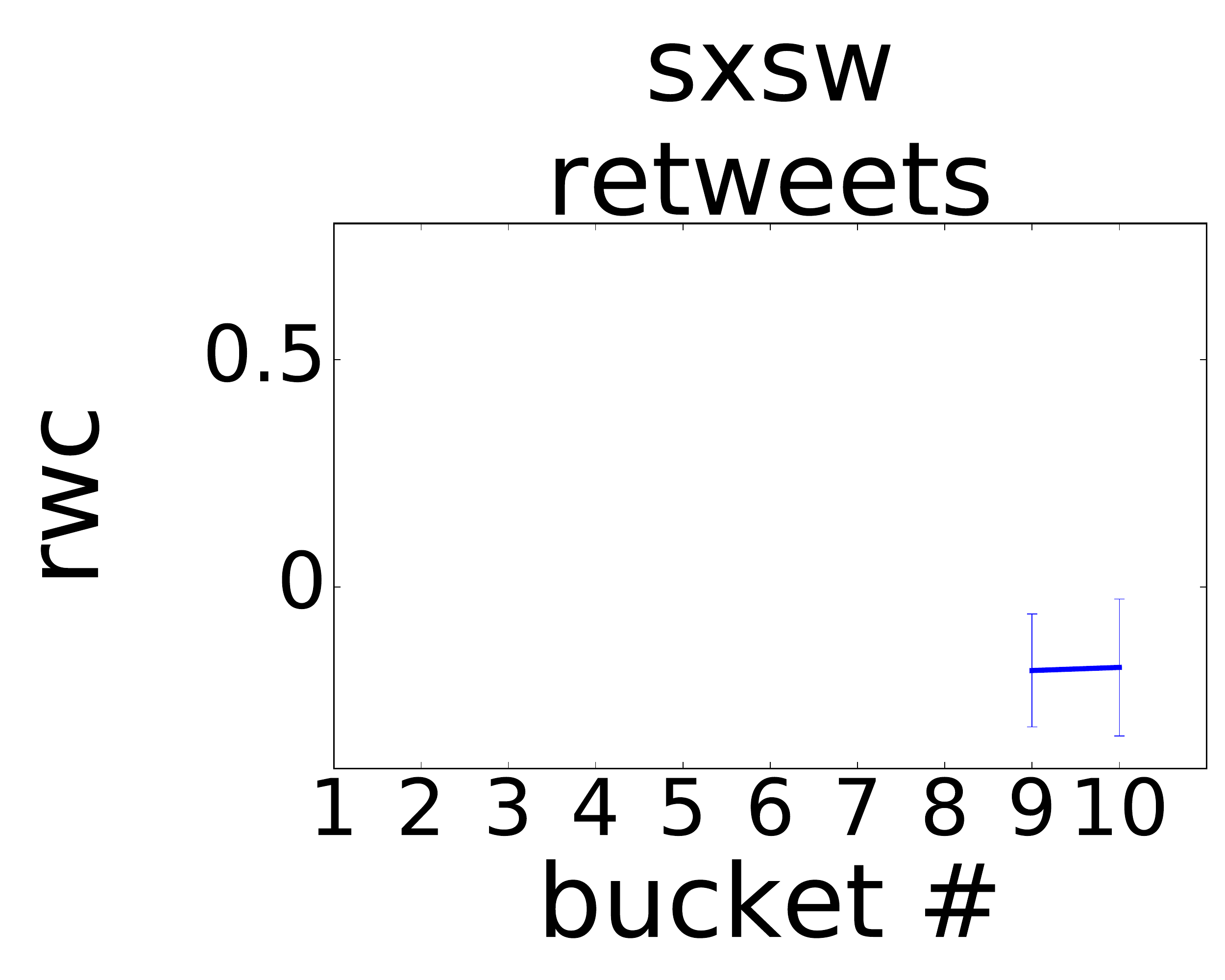}}
\end{minipage}%
\begin{minipage}{.22\linewidth}
\centering
\subfloat{\label{}\includegraphics[width=\textwidth]{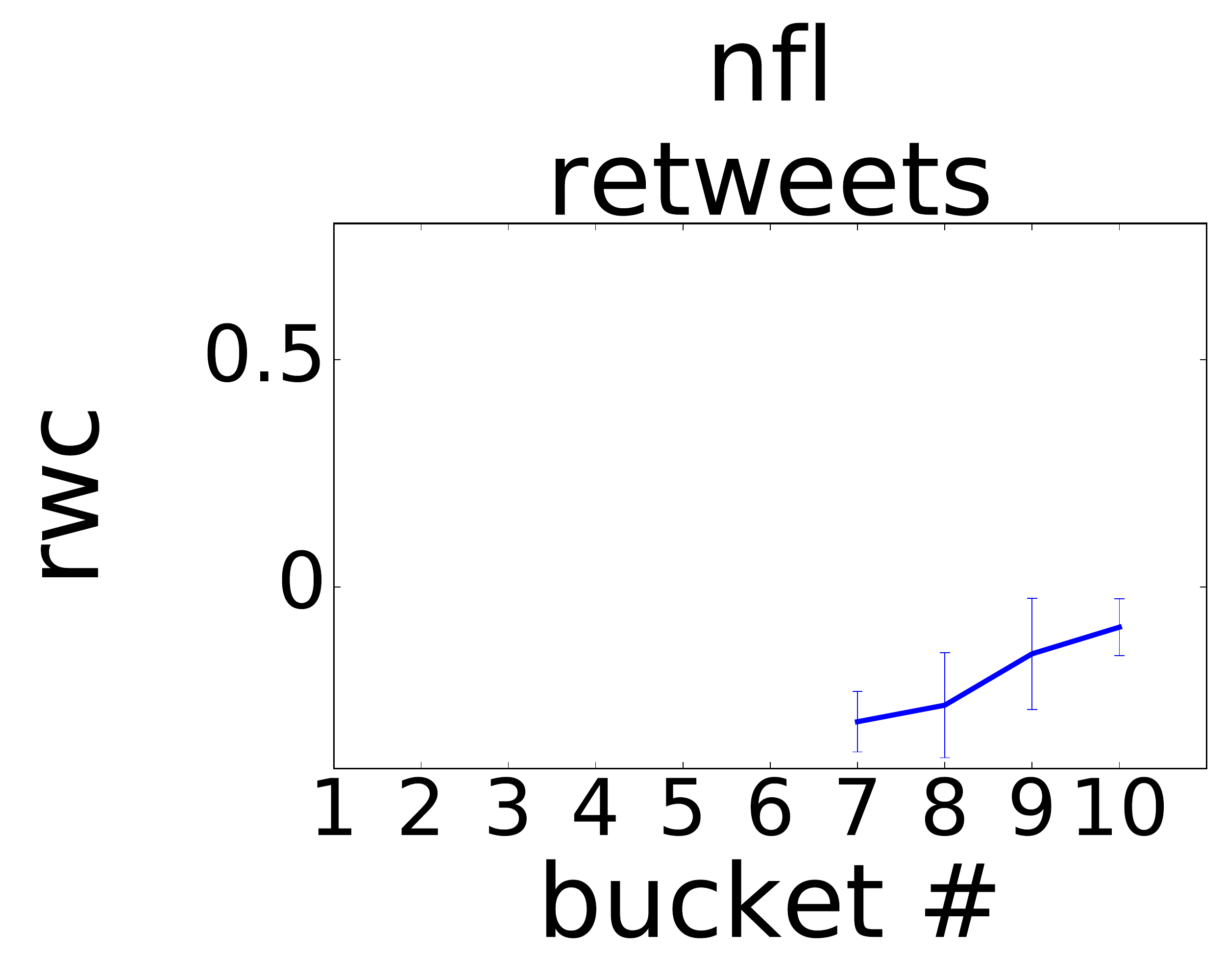}}
\end{minipage}%
\caption{Non-controversial topics: RWC score as a function of the activity in the retweet network. In this case, increase in interest does not affect the controversy score of the networks.}
\label{fig:noncontrrwc}
\end{figure*}

For comparison, we perform measurements over a set of non-controversial topics, defined by the hashtags {\it \#ff}, standing for `Follow Friday', used every Friday by users to recommend interesting accounts to follow; {\it \#nfl}, used to discuss about American football; {\it \#sxsw}, used to comment on the {\it South-by-South-West} conference; {\it \#tbt}, standing for `Throwback Thursday', used every Thursday by users to share memories (news, pictures, stories) from the past.
We find that \rwc\ remains in low value ranges, even as the volume of activity spikes (Figure~\ref{fig:noncontrrwc}).

\begin{figure}[tb]
\begin{center}
\includegraphics[width=\columnwidth]{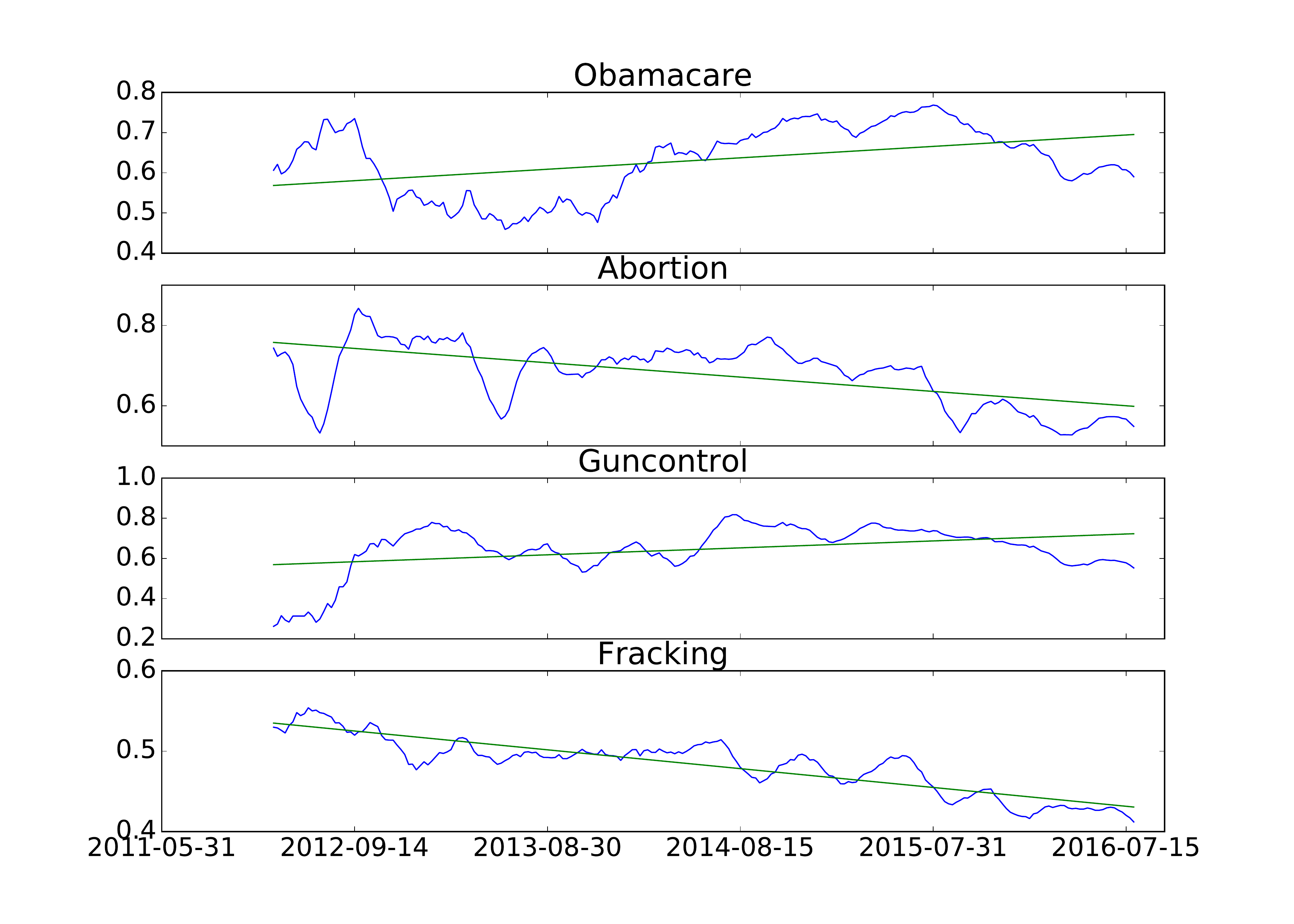}
\caption{Long-term trends of RWC (controversy) score in our dataset. The plots show no consistent trend among the controversial topics under consideration.}
\label{fig:time-trends}
\end{center}
\end{figure}

\subsection{Long-term Evolution of Polarization}
Let us now focus on the long-term evolution of \rwc.
A common point of view holds that social media is aggravating the polarization of society and exacerbating the divisions in it~\cite{benkler2006wealth}.
At the same time, the political debate itself (in U.S.) has become more polarized in recent years~\cite{andris2015rise}.
However, we do not find supportive evidence for this argument in our analysis.

Figure~\ref{fig:time-trends} shows the long-term trends of the RWC measure for the four topics.
The trend is downwards for `abortion' and `fracking' and slightly upwards for `obamacare' and `gun control'.
One could argue that the latter topics are more politically linked to the current administration in U.S., and for this reason have received increasing attention with the approaching elections.
However, the only safe conclusion that can be drawn from this dataset is that there is no clear signal.
The figure suggests that social media, and in particular Twitter, are better suited at capturing the `twitch' response of the public to events and news.
In addition, while our dataset spans a quite long time range for typical social media studies, it is still shorter than ones used typically in social science (coming from, e.g., census, polls, congress votes).
This limit is intrinsic to the tool, given that social media have risen in popularity only relatively recently (Twitter is 10 years old).

\section{Conclusions and Future Work}
\label{sec:conclusions}

We analyzed four controversial topics of discussion on Twitter for a period of five years.
By examining their endorsement networks, we found that spikes in collective attention correspond to an increase in the controversy of the discussion.
However, while instantaneous temporary increase in controversy happens in relation to external events, we did not find evidence of long term increase in polarization.

In future work, we plan to extend our analysis to include other 
network-structure and content-based measures. 
Equally of interest is whether the observations made in this study translate to other social media beside Twitter, for instance, Facebook or Reddit.
Finally, while we did not find any consistent long-term trend in the polarization of the discussions, it is worth continuing this line of investigation, as the effects of increased polarization might not be easily discoverable from social-media analysis alone.

\section*{Acknowledgements.}
This work was supported by the Academy of Finland project ``Nestor'' (286211) and the EC H2020 RIA project ``SoBigData'' (654024).

%References and End of Paper
%These lines must be placed at the end of your paper
\bibliographystyle{aaai}
\bibliography{mathioudakis-icwsm-244}

\begin{thebibliography}{}

\bibitem[\protect\citeauthoryear{Adamic and Glance}{2005}]{adamic2005political}
Adamic, L.~A., and Glance, N.
\newblock 2005.
\newblock The political blogosphere and the 2004 us election: divided they
  blog.
\newblock In {\em LinkKDD},  36--43.

\bibitem[\protect\citeauthoryear{Andris \bgroup et al\mbox.\egroup
  }{2015}]{andris2015rise}
Andris, C.; Lee, D.; Hamilton, M.~J.; Martino, M.; Gunning, C.~E.; and Selden,
  J.~A.
\newblock 2015.
\newblock The rise of partisanship and super-cooperators in the us house of
  representatives.
\newblock {\em PloS one} 10(4):e0123507.

\bibitem[\protect\citeauthoryear{Benkler}{2006}]{benkler2006wealth}
Benkler, Y.
\newblock 2006.
\newblock {\em The wealth of networks: How social production transforms markets
  and freedom}.
\newblock Yale University Press.

\bibitem[\protect\citeauthoryear{Conover \bgroup et al\mbox.\egroup
  }{2011}]{conover2011political}
Conover, M.; Ratkiewicz, J.; Francisco, M.; Gon{\c{c}}alves, B.; Menczer, F.;
  and Flammini, A.
\newblock 2011.
\newblock {Political Polarization on Twitter}.
\newblock In {\em ICWSM}.

\bibitem[\protect\citeauthoryear{{De Francisci Morales}, Gionis, and
  Lucchese}{2012}]{deFrancisciMorales2012trex}
{De Francisci Morales}, G.; Gionis, A.; and Lucchese, C.
\newblock 2012.
\newblock {From Chatter to Headlines: Harnessing the Real-Time Web for
  Personalized News Recommendation}.
\newblock In {\em WSDM},  153--162.

\bibitem[\protect\citeauthoryear{Garimella \bgroup et al\mbox.\egroup
  }{2016a}]{garimella2016exploring}
Garimella, K.; De~Francisci~Morales, G.; Gionis, A.; and Mathioudakis, M.
\newblock 2016a.
\newblock {Exploring Controversy in Twitter}.
\newblock In {\em CSCW [demo]}.

\bibitem[\protect\citeauthoryear{Garimella \bgroup et al\mbox.\egroup
  }{2016b}]{garimella2016quantifying}
Garimella, K.; De~Francisci~Morales, G.; Gionis, A.; and Mathioudakis, M.
\newblock 2016b.
\newblock {Quantifying Controversy in Social Media}.
\newblock In {\em WSDM},  33--42.

\bibitem[\protect\citeauthoryear{Garimella \bgroup et al\mbox.\egroup
  }{2017}]{garimella2017connecting}
Garimella, K.; De~Francisci~Morales, G.; Gionis, A.; and Mathioudakis, M.
\newblock 2017.
\newblock {Reducing Controversy by Connecting Opposing Views}.
\newblock In {\em WSDM}.

\bibitem[\protect\citeauthoryear{Karypis and Kumar}{1995}]{karypis1995metis}
Karypis, G., and Kumar, V.
\newblock 1995.
\newblock {METIS - Unstructured Graph Partitioning and Sparse Matrix Ordering
  System}.

\bibitem[\protect\citeauthoryear{Lehmann \bgroup et al\mbox.\egroup
  }{2012}]{Lehmann2012}
Lehmann, J.; Gon\c{c}alves, B.; Ramasco, J.~J.; and Cattuto, C.
\newblock 2012.
\newblock {Dynamical Classes of Collective Attention in Twitter}.
\newblock In {\em WWW},  251--260.

\bibitem[\protect\citeauthoryear{Lu, Caverlee, and Niu}{2015}]{lu2015biaswatch}
Lu, H.; Caverlee, J.; and Niu, W.
\newblock 2015.
\newblock {BiasWatch: A Lightweight System for Discovering and Tracking
  Topic-Sensitive Opinion Bias in Social Media}.
\newblock In {\em CIKM},  213--222.

\bibitem[\protect\citeauthoryear{Mathioudakis and
  Koudas}{2010}]{mathioudakis2010twittermonitor}
Mathioudakis, M., and Koudas, N.
\newblock 2010.
\newblock {TwitterMonitor: Trend Detection over the Twitter Stream}.
\newblock In {\em SIGMOD},  1155--1158.

\bibitem[\protect\citeauthoryear{Mejova \bgroup et al\mbox.\egroup
  }{2014}]{mejova2014controversy}
Mejova, Y.; Zhang, A.~X.; Diakopoulos, N.; and Castillo, C.
\newblock 2014.
\newblock Controversy and sentiment in online news.
\newblock {\em Symposium on Computation + Journalism}.

\bibitem[\protect\citeauthoryear{Morales \bgroup et al\mbox.\egroup
  }{2015}]{morales2015measuring}
Morales, A.; Borondo, J.; Losada, J.; and Benito, R.
\newblock 2015.
\newblock {Measuring political polarization: Twitter shows the two sides of
  Venezuela}.
\newblock {\em Chaos} 25(3).

\bibitem[\protect\citeauthoryear{Romero, Uzzi, and
  Kleinberg}{2016}]{romero2016social}
Romero, D.~M.; Uzzi, B.; and Kleinberg, J.
\newblock 2016.
\newblock Social networks under stress.
\newblock In {\em WWW},  9--20.

\end{thebibliography}
\end{document}